\documentclass[twocolumn]{jpsj3}

\usepackage{amsmath,amssymb}
\usepackage{graphicx}
\usepackage{xcolor}
\usepackage{color}

\usepackage[final]{changes}

\definechangesauthor[color=red]{AP}
\definechangesauthor[color=blue]{GK}

\usepackage{gensymb}

\usepackage{amsmath,amssymb}
\usepackage{graphicx}
\usepackage{xcolor}
\usepackage{ulem}

\begin{document}

\title{\textbf{Fermi Surfaces in the Antiferromagnetic, Paramagnetic and Polarized Paramagnetic States of CeRh$_{2}$Si$_{2}$ Compared with Quantum Oscillation Experiments}}

%\author{A. Pourret}
%\email[E-mail, ]{alexandre.pourret@cea.fr}
%\affiliation{University Grenoble Alpes, INAC-PHELIQS, F-38000 Grenoble, France}
%\affiliation{CEA, INAC-PHELIQS, F-38000 Grenoble, France}
%\author{M.-T. Suzuki}
%\affiliation{RIKEN Center for Emergent Matter Science, Hirosawa 2-1, Wako, Saitama 351-0198, Japan}
%\author{A. Palacio Morales}
%\affiliation{University Grenoble Alpes, INAC-PHELIQS, F-38000 Grenoble, France}
%\affiliation{CEA, INAC-PHELIQS, F-38000 Grenoble, France}
%\author{G. Seyfarth}
%\affiliation{University Grenoble Alpes, LNCMI, F-38042 Grenoble Cedex 9, France}
%\affiliation{CNRS, Laboratoire National des Champs Magn\'etiques Intenses LNCMI (UJF, UPS, INSA), UPR 3228, F-38042 Grenoble Cedex 9, France}
%\author{G. Knebel}
%\affiliation{University Grenoble Alpes, INAC-PHELIQS, F-38000 Grenoble, France}
%\affiliation{CEA, INAC-PHELIQS, F-38000 Grenoble, France}
%\author{D. Aoki }
%\affiliation{University Grenoble  Alpes, INAC-PHELIQS, F-38000 Grenoble, France}
%\affiliation{CEA, INAC-PHELIQS, F-38000 Grenoble, France}
%\affiliation{Institute for Materials Research, Tohoku University, Oarai, Ibaraki, 311-1313, Japan}
%\author{J. Flouquet}
%\affiliation{University Grenoble Alpes, INAC-PHELIQS, F-38000 Grenoble, France}
%\affiliation{CEA, INAC-PHELIQS, F-38000 Grenoble, France}
%\date{\today }

\author{
Alexandre~{Pourret}$^{1,}$\thanks{E-mail address, alexandre.pourret@cea.fr},
Michi-To~{Suzuki}$^{2}$,
Alexandra~{Palaccio Morales}$^{1}$,
Gabriel~{Seyfarth}$^{3,4}$,
Georg~{Knebel}$^{1}$,
Dai~{Aoki}$^{1,5}$,
and
Jacques~{Flouquet}$^{1}$%\\
}

\inst{%
$^1$Universit\'e Grenoble Alpes, CEA, INAC, PHELIQS, F-38000 Grenoble, France\\
$^2$RIKEN Center for Emergent Matter Science, Hirosawa 2-1, Wako, Saitama 351-0198, Japan\\
$^3$Universit\'e Grenoble Alpes, LNCMI, F-38042 Grenoble Cedex 9, France\\
$^4$CNRS, Laboratoire National des Champs Magn\'etiques Intenses LNCMI (UGA, UPS, INSA), UPR 3228, F-38042 Grenoble Cedex 9, France\\
$^5$Institute for Materials Research, Tohoku University, Oarai, Ibaraki 311-1313, Japan

}

\date{\today }

%\begin{abstract}
\abst{The large quantum oscillations observed in the thermoelectric power in the antiferromagnetic (AF) state of the heavy-fermion compound CeRh$_2$Si$_2$ disappear suddenly when entering in the polarized paramagnetic (PPM) state at $H_{c}\sim 26.5$~T, indicating an abrupt reconstruction of the Fermi surface. The electronic band structure was [LDA+$U$] for the AF state taking the correct magnetic structure into account, for the PPM state, and for the paramagnetic state (PM). Different Fermi surfaces were obtained for the AF, PM, and PPM states. Due to band folding, a large number of branches was expected and observed in the AF state. The LDA+$U$ calculation was compared with the previous LDA calculations. Furthermore, we compared both calculations with previously published de Haas-van Alphen experiments. The better agreement with the LDA approach suggests that above the critical pressure $p_c$ CeRh$_2$Si$_2$ enters in a mixed-valence state. In the PPM state under a high magnetic field, the $4f$ contribution at the Fermi level $E_F$ drops significantly compared with that in the PM state, and the $4f$ electrons contribute only weakly to the Fermi surface in our approach.
}
%\end{abstract}

%\pacs{}

%\keywords{}

\maketitle

\section{Introduction}

Heavy-fermion metals are characterized by the formation of heavy quasiparticles below a characteristic energy scale due to the interplay of the $f$ electrons with the light conduction electrons. This gives rise to the formation of a Kondo singlet and, in the case of full compensation of the $f$ moment by the conduction electrons, a paramagnetic (PM) ground state. However, in the case of a magnetically ordered ground state, the interplay between the localization of the quasiparticles, the magnetism, and the Fermi surface topology in heavy-fermion metals is still under debate.\cite{Lohneysen2007} In a classical $4f$ antiferromagnet, the conventional view is that on entering the antiferromagnetic (AF) state, a Fermi surface reconstruction will occur owing to the band folding induced by the reduction of the Brillouin zone associated with the AF ordering. Additionally, in heavy-fermion systems close to a magnetic instability induced by pressure ($P$) or magnetic field ($H$), the drastic change in the dependence on the localization of the $4f$ electrons has been discussed on the basis of the concept of local criticality with the change from an AF ordered small Fermi surface to a large Fermi surface in the PM state.\cite{Si2010} In the magnetically ordered state, the $4f$ electrons do not participate in the Fermi surface, while the heavy quasiparticles in the PM phase should contribute. Among these systems, the Ising-like antiferromagnet CeRh$_{2}$Si$_{2}$ is particularly interesting as its magnetic structure is well known and commensurable with the crystal structure.\cite{Kawarazaki2000} This allows the band structure in the AF phase to be calculated while explicitly taking into account the magnetic ordering and the change in the Brillouin zone, which have been omitted in all previous contributions to our knowledge.

\begin{figure}
	\begin{center}
	\includegraphics[width=8cm]{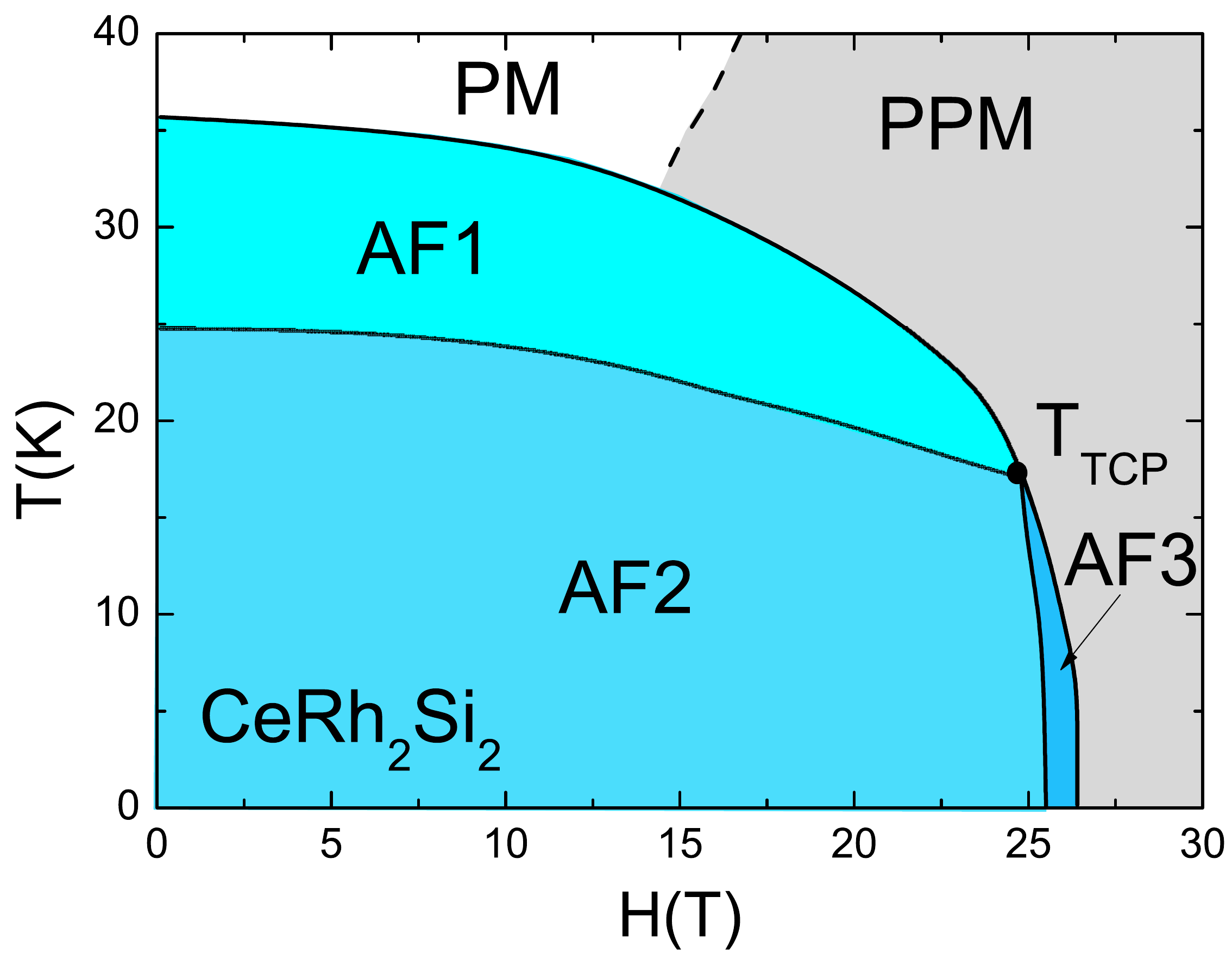}
	\end{center}
	\caption{\label{Fig1} (Color online) $T-H$ phase diagram of CeRh$_{2}$Si$_{2}$ with the presence of three different AF states at low temperature, from Ref. \citen{PalacioMorales2015}. The dashed line (taken from Ref.~\citen{Knafo2010}) characterizes a crossover between the low-field PM regime, where AF correlations dominate, and the high-field PPM regime.}
\end{figure}

CeRh$_2$Si$_2$ crystallizes in the ThCr$_2$Si$_2$-type structure with space group I4/mmm. Its magnetic phase diagram is shown schematically in Fig.~\ref{Fig1}. On cooling at zero field and ambient pressure, CeRh$_{2}$Si$_{2}$ undergoes a first AF transition to the AF1 phase at $T_{N1}=36$~K, characterized by the wave vector $\mathbf{q_{1}}=(\frac{1}{2}$,~$\frac{1}{2}$,~0). The magnetic structure changes at $T_{N2}=25$~K, and in the low temperature AF2 phase an additional wave vector ${\bf q_{2}}=(\frac{1}{2}$,~$\frac{1}{2}$,~$\frac{1}{2}$) appears, forming a 4-{\bf q} superposed structure.\cite{Kawarazaki2000}
The sublattice magnetization $M_{0}$ per Ce atom at 4.2~K is approximately 1.5~$\mu_{B}$.\cite{Kawarazaki2000} The Sommerfeld coefficient $\gamma=\frac{C}{T}$ extrapolated to $T\rightarrow0$~K is only 23~mJmol$^{-1}$K$^{-2}$ (see Ref.~\citen{Graf1998}). However, the extrapolation of $\frac{C}{T}$ to $T \rightarrow 0$~K  from the PM phase by taking the entropy balance into account gives $\gamma \approx 300$~mJmol$^{-1}$K$^{-2}$. This clearly indicates that CeRh$_2$Si$_2$ is a heavy-fermion compound in its PM phase.

Under a magnetic field along the $c$ axis, the AF order is suppressed and the system enters a polarized paramagnetic (PPM) regime. For $T<18$~K a cascade of two first-order transitions appears, AF2 $\rightarrow$ AF3 $\rightarrow$ PPM, with two critical metamagnetic fields $H_{2-3}=25.7$~T and $H_{c}=26$~T associated with two quasi-identical magnetization jumps of 0.75~$\mu_B$/Ce (see Fig.~\ref{Fig1}).\cite{Settai1997,Abe1997} As the AF3 phase exists only in a narrow magnetic field range and its magnetic structure has not yet been determined, the main focus later in this paper will be on the differences between the AF2 and PPM phases.

Up to now, Fermi surface studies at ambient pressure have been restricted to the AF phase, i.e $H < H_{c}$.\cite{Araki2001} Our recent high magnetic field thermoelectric power (TEP) experiment showed that an important reconstruction of the Fermi surface appears at $H_c$.\cite{PalacioMorales2015} Various macroscopic measurements suggest that at high magnetic field above $H_{c}$, the PPM ground state differs from the PM state.\cite{Knafo2010} No band structure calculation has been reported for the AF and PPM phases in this system.\cite{Settai1997,Abe1997, Knafo2010} The pressure dependence of quantum oscillations suggests a Fermi surface reconstruction at $p_c$ ($\sim 1$~GPa), where the AF order is suppressed and the system enters a PM state above $p_{c}$. \cite{Araki2001}

In a previous article, we presented TEP experiments on CeRh$_2$Si$_2$ in the AF state at ambient pressure under a high magnetic field and under pressure.\cite{PalacioMorales2015} We were able to observe quantum oscillations in the AF state at ambient pressure, which abruptly disappeared when entering the PPM state above $H_c$. We also presented the electronic density of states (DOS) in the AF regime by taking the real magnetic structure into account as well as that in the PM regime, and we showed that at $P = 0$ in the AF phase the 4$f$ contribution at the Fermi level ($E_F$) is weak, while it is the main contribution in the PM domain.

In the present article, we present the calculated Fermi surfaces and the DOS in the different states of the phase diagram shown in Fig.~\ref{Fig1}: the high temperature PM state, the AF state, and the PPM state under high magnetic field. We compare them with the results of a TEP quantum oscillation experiment.
The aim of this study is to provide a consistent picture for changes in the Fermi surface, which are expected from our TEP quantum oscillation data, by performing first-principles electronic structure calculations using the Local Density Approximation+$U$ (LDA+$U$) method. These calculations lead to different topologies of the Fermi surfaces for the PM, PPM, and AF states. The calculations show that in the PM state, the $f$ states are itinerant and located around the Fermi level, while the PPM state has a Fermi surface that appears as if the $f$ states are localized by the large magnetic splitting of the lower and upper Hubbard bands. In the AF state the $f$ states are located well below the Fermi level. Furthermore, the AF state produces a large number of Fermi surfaces owing to the magnetic band folding, leading to a large number of de Haas-van Alphen (dHvA) branches as experimentally observed here.

\section{Experimental Results}
\begin{figure}[h!]
	\begin{center}
	\includegraphics[width=8.5cm]{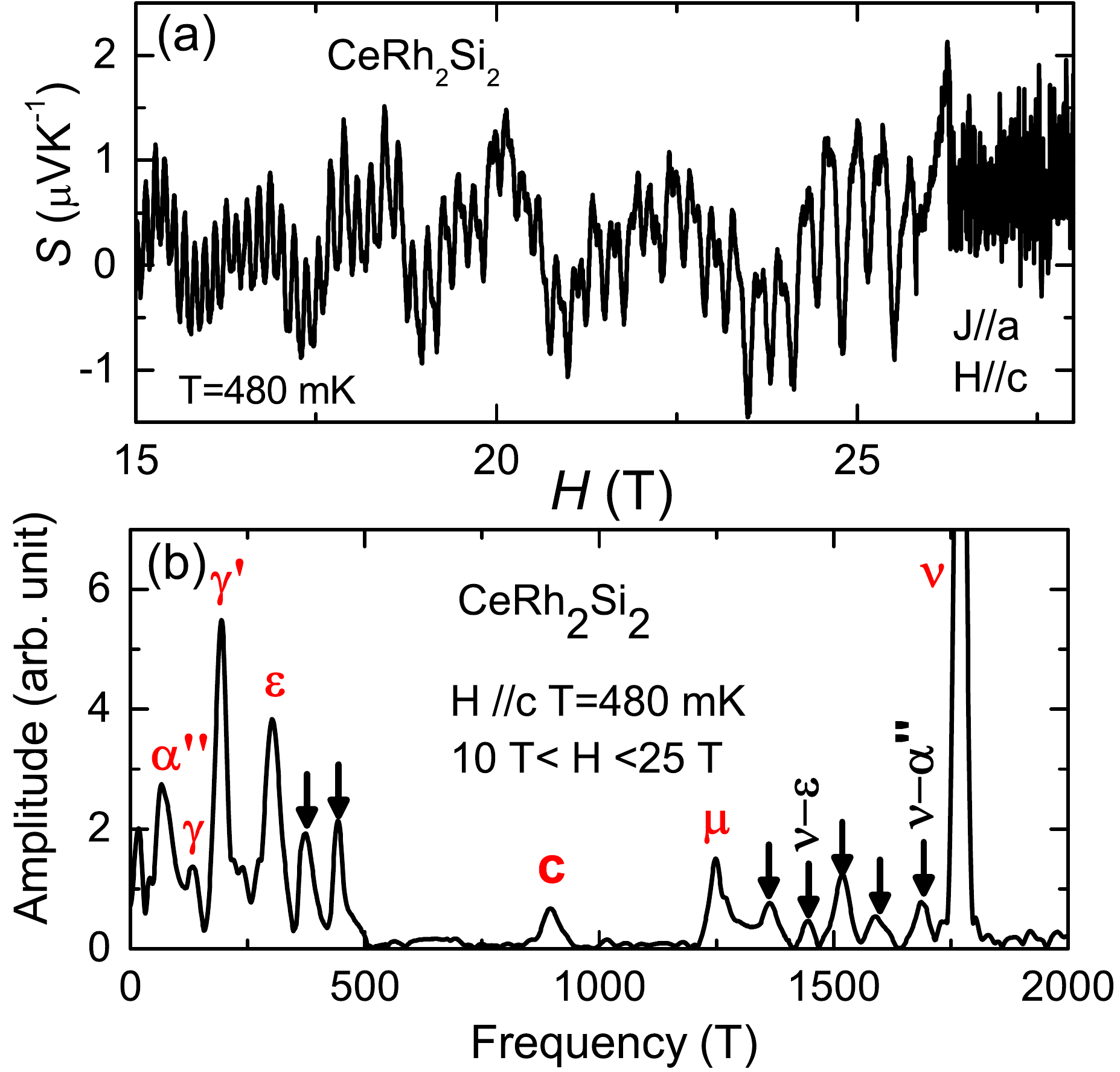}
	\end{center}
	\caption{\label{Fig2} (Color online) (a) TEP, $S(H)$, as a function of magnetic field at $T$=480~mK for the transverse configuration obtained from previous measurements.\cite{PalacioMorales2015} Quantum oscillations disappear when entering the PPM phase at $H_{c}=26$~T. (b) Fast Fourier transformation of $S(H)$ at $T=480$~mK in the field range from 10 to 26~T. The labels of the Fermi surface branches follow those of Ref.~\citen{Araki2001}. Arrows indicate branches that have not been observed previously.}
\end{figure}

The TEP as a function of magnetic field applied along the $c$ axis was previously reported in Fig.~6 of Ref.~\citen{PalacioMorales2015}. It was measured up to a high magnetic field of 34~T at LNCMI Grenoble, and measurements up to 16~T were also performed in a superconducting magnet down to 180~mK. In Fig.~\ref{Fig2} we plot the magnetic field dependence of the TEP at $T=480$~mK with $J\parallel a$ and $H\parallel c$. As already mentioned in Ref.~\citen{PalacioMorales2015}, large quantum oscillations appear up to $H_{2,3}$ while no quantum oscillation signal is detected above $H_{c}$. To obtain information on the Fermi surface properties, we performed a fast Fourier transform (FFT) analysis of the signal obtained in the AF2 state.

\begin{figure}[h!]
	\begin{center}
	\includegraphics[width=8.5cm]{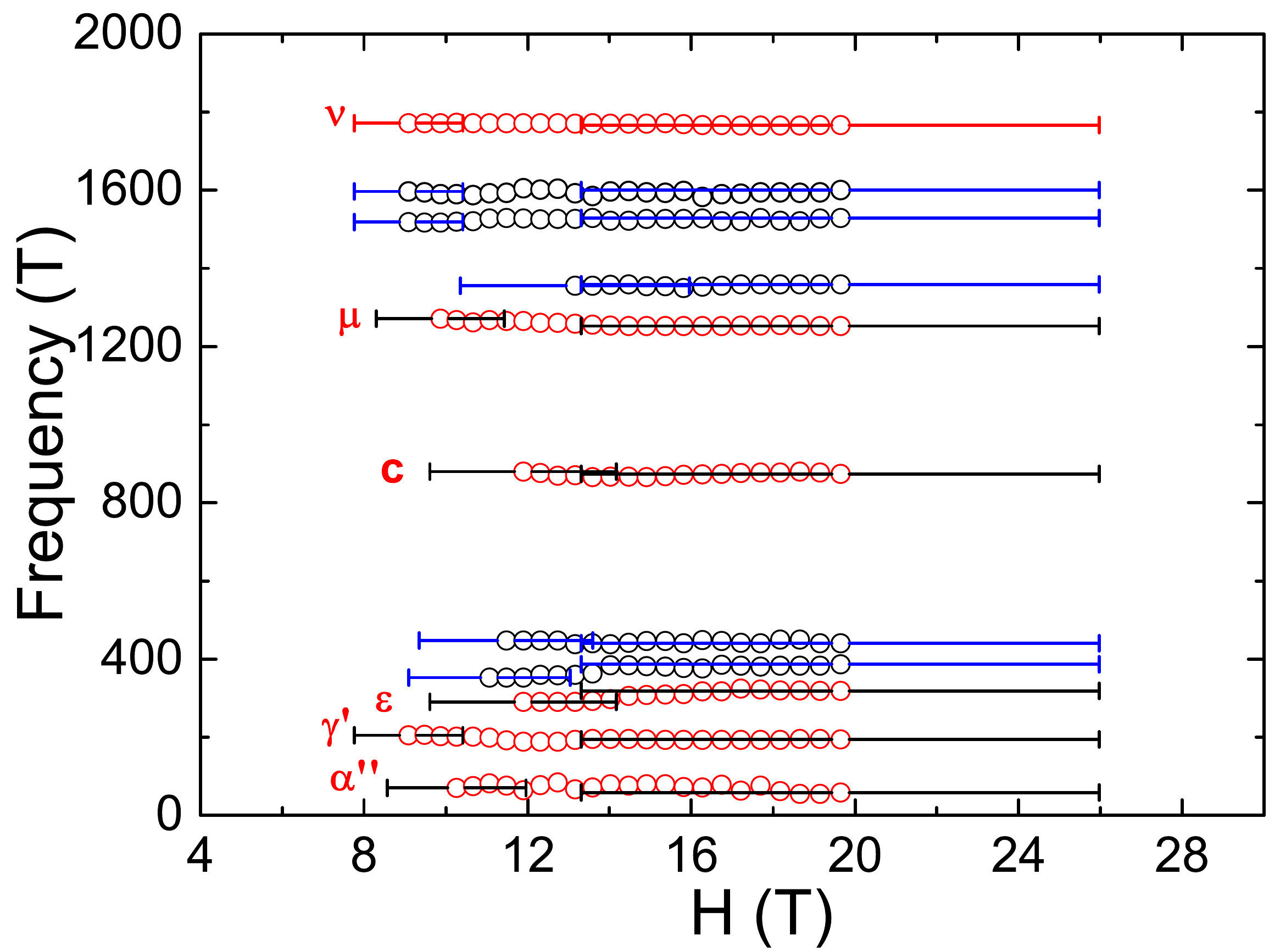}
	\end{center}
	\caption{\label{Fig3} (Color online) Magnetic field dependence of the TEP quantum oscillation frequencies. The FFT was performed on a constant inverse field window ($\frac{1}{H_{\rm min}}-\frac{1}{H_{\rm max}}=0.03~$T$^{-1}$) that was shifted with the field. The horizontal bars give the field window for the FFT analysis.}
\end{figure}

Figure \ref{Fig2}(b) shows a typical FFT spectrum for TEP quantum oscillations in CeRh$_2$Si$_2$ with the field along the $c$ axis in the field range of $10-25$~T. Fourteen frequencies were observed in this direction. In addition to the main branches ($\alpha"$, $\gamma'$, $\gamma$, $\epsilon$, c, $\mu$, and $\nu$) previously identified in dHvA experiments,\cite{Araki2001} seven additional frequencies (indicated by black vertical arrows) were observed.  All frequencies are reported in Table~\ref{Table}. Two of these frequencies may have been attributable to linear combinations of two fundamental frequencies, $\nu-\alpha"$ and $\nu-\epsilon$, while the five others seem to correspond to previously undetected Fermi surface branches. \cite{note} In order to investigate the field dependence of the observed quantum oscillation frequencies, we performed an FFT analysis over a sliding window of width $\frac{1}{H_{\rm min}}-\frac{1}{H_{\rm max}}=0.03$~T$^{-1}$.\cite{oscillations}  As shown in Fig.~\ref{Fig3} for the main frequencies, none of the observed frequencies show a magnetic field dependence until their sudden disappearance above $H_c$.  However, the maximal effective field $1/H_{\rm eff} = 1/2(1/H_{\rm min} + 1/H_{\rm max})$ is below $H_c$. This constant evolution of the frequencies below $H_c$ contrasts with the sudden change in the Fermi surface at the first-order transition at $H_c$.

\begin{figure}[h!]
	\begin{center}
	\includegraphics[width=8.5cm]{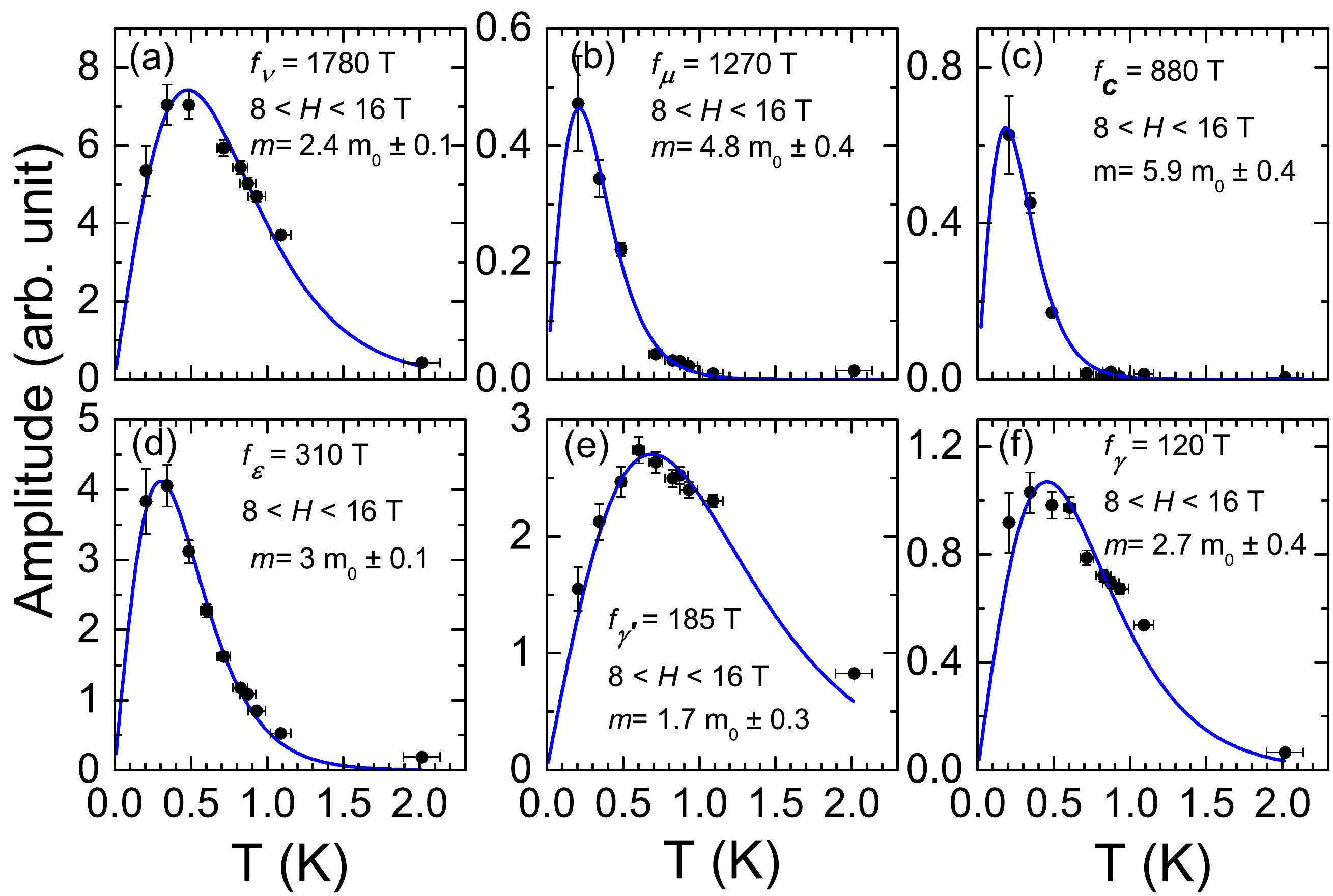}
	\end{center}
\caption{\label{Fig4} (Color online) Temperature dependence of
the FFT amplitudes for the main branches. The blue lines are least-squares fits to Eq.~(\ref{PV}). The error bar of the FFT amplitude was estimated from the maximal error in the thermal gradient. The accordingly obtained effective masses are also indicated.}
\end{figure}

\begin{figure}[h!]
	\begin{center}
	\includegraphics[width=1.0\linewidth]{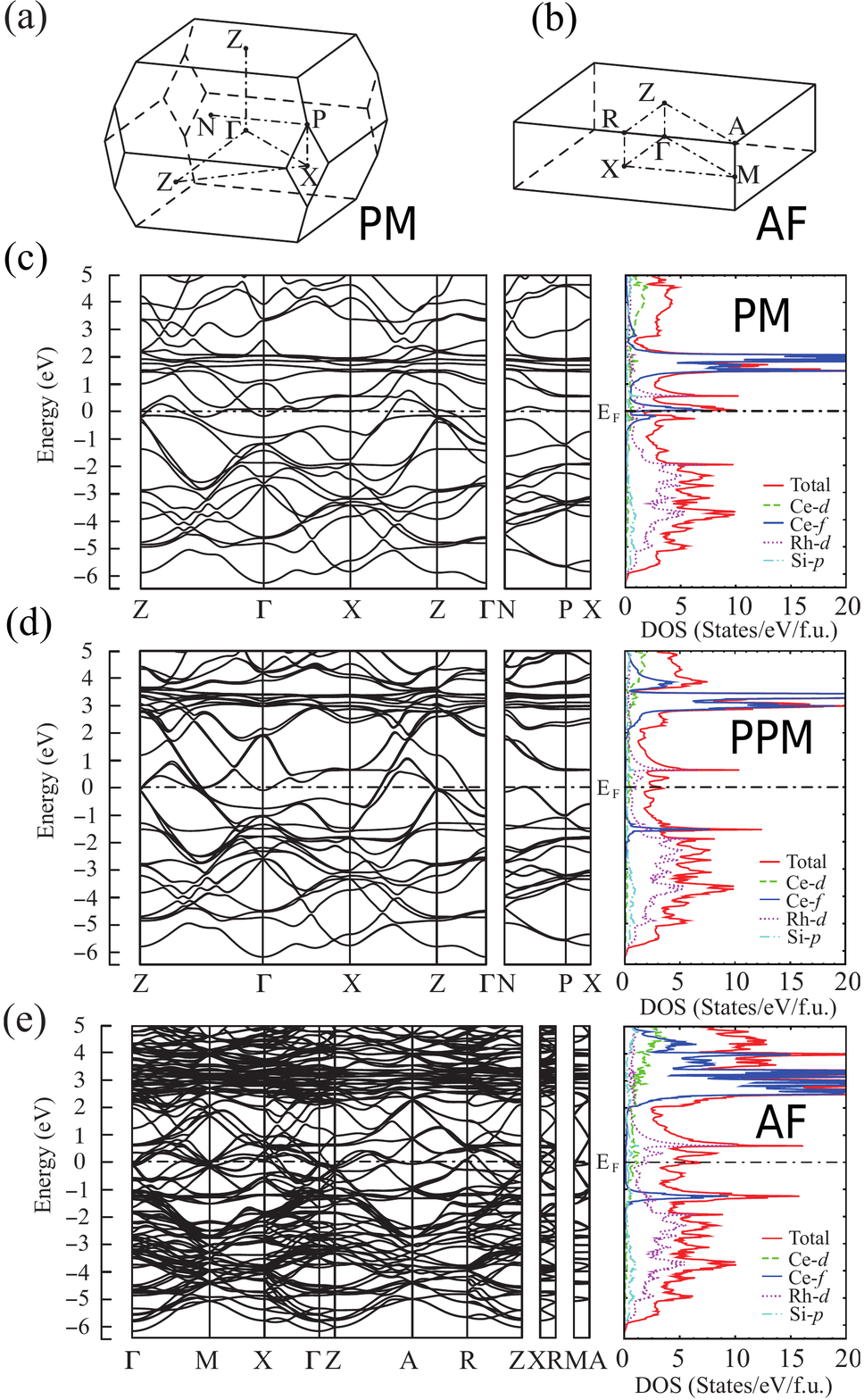}
	\end{center}
	\caption{\label{Fig5} (Color online) (a) Brillouin zone of the PM phase of CeRh$_2$Si$_2$. (b) Brillouin zone of the AF phase. (c) Band structure and DOS of the PM state of CeRh$_2$Si$_2$. (d) Band structure and DOS of the PPM state. (e) Band structure and DOS of the AF state.
\label{Fig,BandDOS}}
\end{figure}

To determine the cyclotron masses $m_{c}$ of each branch, we analyzed the temperature dependence of the amplitudes of the FFT spectra in the field range from 8 to 16~T for each branch following the theoretical prediction for the amplitude of the quantum oscillations of the TEP,\cite{Young1973,Fletcher1981,Pantsulaya1989} where large oscillations due to the energy dependence of the relaxation time of the electrons under a high magnetic field can be observed, which gives the main contribution to the oscillatory part,\cite{Pantsulaya1989}

\begin{equation}
\label{PV}
A(T) \propto \frac{(\alpha pX) \coth(\alpha pX)-1}{\sinh(\alpha p X)}.
\end{equation}
Here $\alpha= 2\pi^2 k_B/e \hbar$, $p$ is the order of the harmonics, and $X=m^*_c T/H_{\rm eff}$.  More details on the determination of the effective mass are given in Ref.~\citen{Palacio2015b}. As the temperature and temperature gradient were not perfectly constant during the field sweep owing to the field dependence of the thermal conductivity, we corrected this by averaging the temperature in the corresponding field window.
Figure~\ref{Fig4} shows the temperature dependence of the FFT amplitudes of the main frequencies. The error bars in the amplitude of the thermoelectric quantum oscillations and in the temperature are indicated by vertical and horizontal lines, respectively.  Owing to the very large amplitude at the frequencies of the FFT spectra compared with the white background noise, the vertical errors are mainly dominated by the error in the applied thermal gradient, reaching an error of around 12\% at the lowest temperature when the thermal gradient is small.  The error in the temperature is directly linked to the applied thermal gradient.
The temperature dependence of the FFT amplitudes is well fitted by the theoretical prediction (Eq.~(\ref{PV})). The position of the maximum of the amplitude of each branch as a function of temperature is given by $T^*\approx \frac{0.11H_{\rm eff}}{pm^*_c }$.
It is obvious that the amplitude is lowest for the lowest temperature for most of the shown frequencies and that the oscillatory and non-oscillatory contributions of the TEP vanish for $T \rightarrow 0$ following the third law of thermodynamics. In Table~I we have listed the effective masses obtained from the analysis of the TEP and the previous dHvA experiments, \cite{Araki2001} and very good agreement was found.

\begin{table*}[h]
\caption{\label{Table}List of quantum oscillation frequencies in the AF2 phase of CeRh$_2$Si$_2$ for $H\parallel c$ obtained from the dHvA measurements of Ref.~\citen{Araki2001}
and from $S(H)$ at $T=468$~mK for the transverse configuration compared with calculated values.\\}
\begin{tabular}{ccccccc}
  %\multicolumn{7}{c}{Frequencies of the $AF2$ phase (T)}\\
  \hline
 %\multicolumn{4}{c}{ branches} & dHvA  [13-16.9]~T (Ref.~\citen{Araki2001}) & $S(H)$ $[8-16]$~T & Theoretical prediction

\multicolumn {1}{c}{ branches} &\multicolumn {2}{c}{dHvA (13--16.9~T)(Ref.~\citen{Araki2001}) }& \multicolumn{2}{c}{$S(H)$ (8--16~T)} & \multicolumn{2}{c}{LDA+$U$} \\
\multicolumn {1}{c}{ } & $F$~(T) & $m^\star$~($m_0$) & $F$~(T) & $m^\star$~($m_0$) &$F$~(T) & $m^\star$~($m_0$)   \\
 \hline
 l & 44 & 0.43 & 44.6 &1.0& 37 &0.52\\
  k& 56 &0.41 &         &       & 55  &0.26 \\
  j& 66 &0.39  &	        &        &       &  \\
 d & 77 &0.26 &        &        & 72  &  0.12 \\
  $\alpha ^{''}$& 81   &  1.8  & 78 &   1.6 & 89 & 0.38  \\
  $\gamma$& 137 & 0.45 & 120 &  2.7    &147& 0.36  \\
  $\gamma'$& 184 &1.4 & 185 &1.7& 174 & 0.25 \\
  &  & &255& 2.4      &  256 &0.56\\
  $\varepsilon$& 327 & 1.9 & 310 &3&  &\\
  & && 377& 4.1  & 362& 1.38\\
  & & & 445 & 5.6 & 448&1.26 \\
  c& 804 &4.9 & 880 &5.9  &881&0.59\\
  $\mu$& 1160 &3.7 & 1270 & 4.8  & 1174&1.43 \\
  & & &1370 & 7.2 &1331 &1.79  \\
  $\nu$-$\varepsilon$    & & &1448 ($\sim$1776-311) & 6.0 ($\sim$3+2.4) &&\\
  & & & 1530 &4.7 &\\
  & & & 1594 &3.6& \\
  $\nu-\alpha^{''}$   & & &1693 ($\sim$1776-78) & 4.2($\sim$1.6+2.4)& \\
  $\nu$ & 1770 & 2.4  & 1780 &2.4  \\
  a & 7560 &6.4 & && 7489&2.26 \\
    \hline
\end{tabular}
\end{table*}

\section{Electronic Structure}
\subsection{Method}

The LDA+$U$ band structure calculations were performed in the framework of the full potential linearized augmented plane wave (FLAPW) method. In the FLAPW method, the scalar relativistic effects are taken into account for all electrons, and the spin-orbit interactions are included self-consistently for all valence electrons in a second variational procedure. The LDA+$U$ method in this study was introduced to add a Hartree-Fock-like effective potential to the LDA, which is accurate for a homogeneous electron gas. Therefore, $U$ is usually considered only for electrons  with localized character ($d$ and $f$), where the large Coulomb repulsion cannot be described in the LDA. Moreover, this method possesses a tendency that the $f$ electron charge distribution becomes anisotropic compared with the LDA calculations. The calculations were carried out under the assumptions of the symmetries expected for the PM, AF, and PPM states.

First, the calculations for the PM state were performed with the LDA+$U$ method assuming time-reversal symmetry. Previously, similar calculations were performed to reproduce the Fermi surface in the PM state of CeRu$_2$Si$_2$~\cite{Suzuki2010} and CeCu$_2$Si$_2$.\cite{Ikeda2015} Second, in order to calculate the PPM state, we extended the PM calculation by assuming a ferromagnetic (FM) alignment along [001]. In the PPM state we expected localized $f$ states, this is the reason why we calculated the FM state for the PPM state, in which the $f$ states do not contribute to the Fermi surfaces. Of course, the PPM state is strictly different from a FM state as it is stable only under a high magnetic field. Finally, the Fermi surfaces of the AF state were investigated by taking into account the correct magnetic alignment in the magnetic unit cell of the 4-{\bf q} magnetic structure.

In the calculations, we included Ce $5p/5d/6s/4f$, and Rh $4d/5s$ Si $3s/3p$ orbitals as valence states and Ce $5s/4d$, Rh $4s/4p$, and Si $2s/2p$ orbitals as semi-core states. As a result, one formula unit cell of CeRh$_2$Si$_2$ contains 80 band electrons.
The AF state contains eight formula units of CeRh$_2$Si$_2$ in the magnetic unit cell, which contains 640 band electrons. The PM state exhibits degeneracy of the energy bands owing to the presence of space inversion and time-reversal symmetry.  The AF state also has energy band degeneracy because of the symmetry of the continuous transformation of the time reversal and space inversion with the body center translation in the magnetic unit cell. We therefore indexed the energy bands by neglecting the degenerated pair of each band.
The lattice parameters determined at 10~K are taken from Ref. \citen{Grier1984} with $a=4.0828$~\AA, $c=10.1705$~\AA, $z=0.3737$ (silicon position within the unit cell), $a_{4Q-AFM}=\sqrt{2}a$, and $c_{4Q-AFM}=2c$. We used these parameters for the AF, PM, and PPM states. The change in the lattice parameters due to thermal expansion or magnetostriction effects was not taken into account and the different DOSs and Fermi surfaces are solely due to changes in the magnetic structure. The radii of muffin-tin spheres, which separate the region using plane waves and atomic bases for the bases, are 1.60~\AA~ for Ce, and 1.17~\AA~for Rh and Si.  We adapted $U=5$~eV on the basis of photoemission experiments on Ce atoms.\cite{Cox1981} For the LDA+$U$ double-counting term, the fully localized limit was adapted. This is an important ingredient of the Fermi surfaces in the PM states. For the calculation of magnetic states, the localized nature of $f$ electrons is a consequence of the large $U$ values introduced as the Hartree-Fock-like potential in the LDA+$U$ method with the magnetic order. \cite{Suzuki2010} Note that different $U$ values do not affect the essential features of the Fermi surfaces unless the $U$ value is too small to induce splitting between the lower and upper Hubbard bands.

\begin{figure}[h!]
	\begin{center}
	\includegraphics[width=1.0\linewidth]{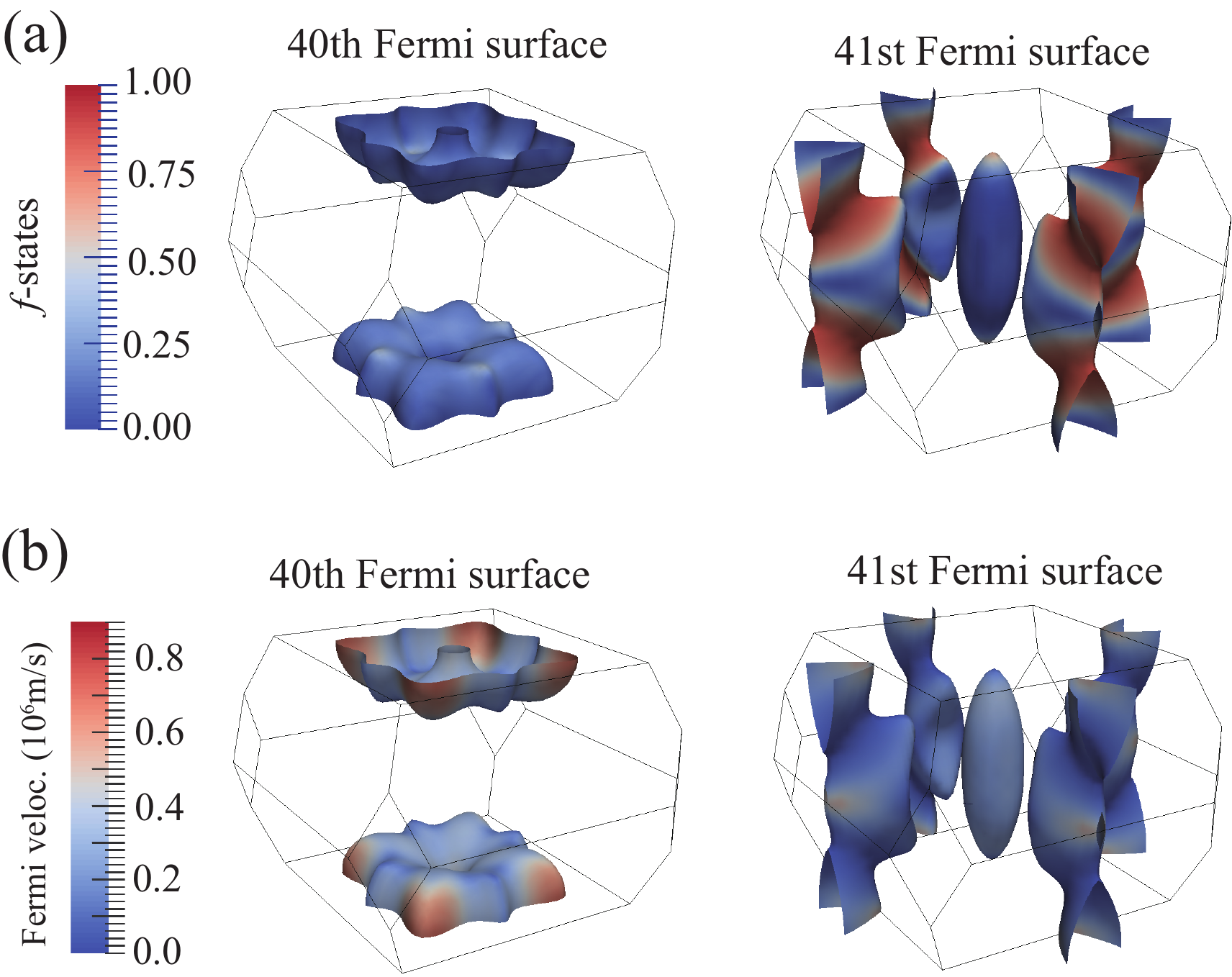}
	\end{center}
	\caption{\label{Fig6} (Color online) Calculated Fermi surfaces
 in the PM states showing (a) $f$ state contribution and
 (b) Fermi velocity.}
\end{figure}

\begin{figure}[h!]
	\begin{center}
	 \includegraphics[width=1.0\linewidth]{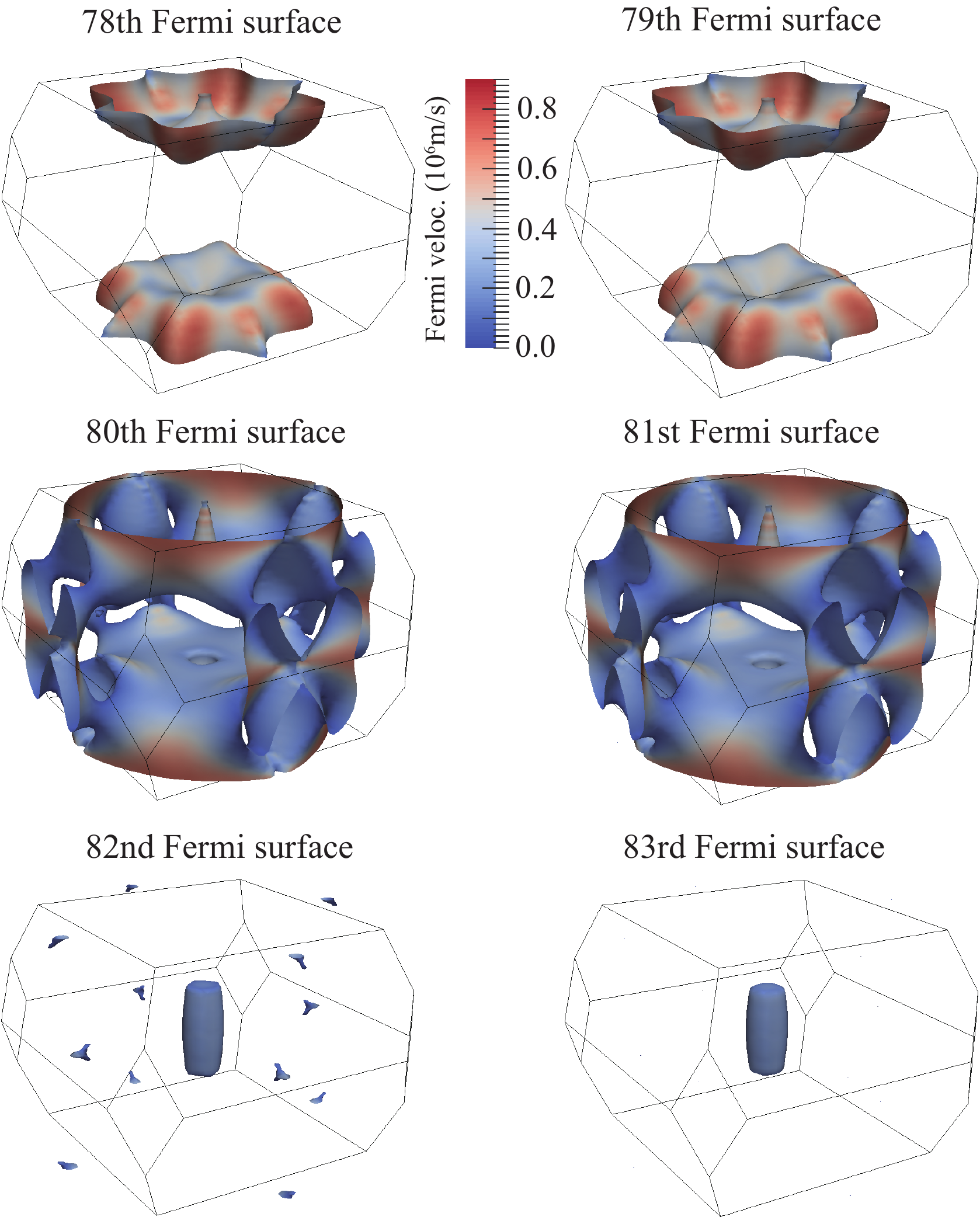}
	\end{center}
	\caption{\label{Fig7} (Color online) Calculated Fermi surfaces
 and Fermi velocities in the PPM state. The numerical range of the
 color legend for the Fermi velocity is the same as that in Fig.~\ref{Fig6}(b).}
\end{figure}

\begin{figure}[h!]
	\begin{center}
	\includegraphics[width=1.0\linewidth]{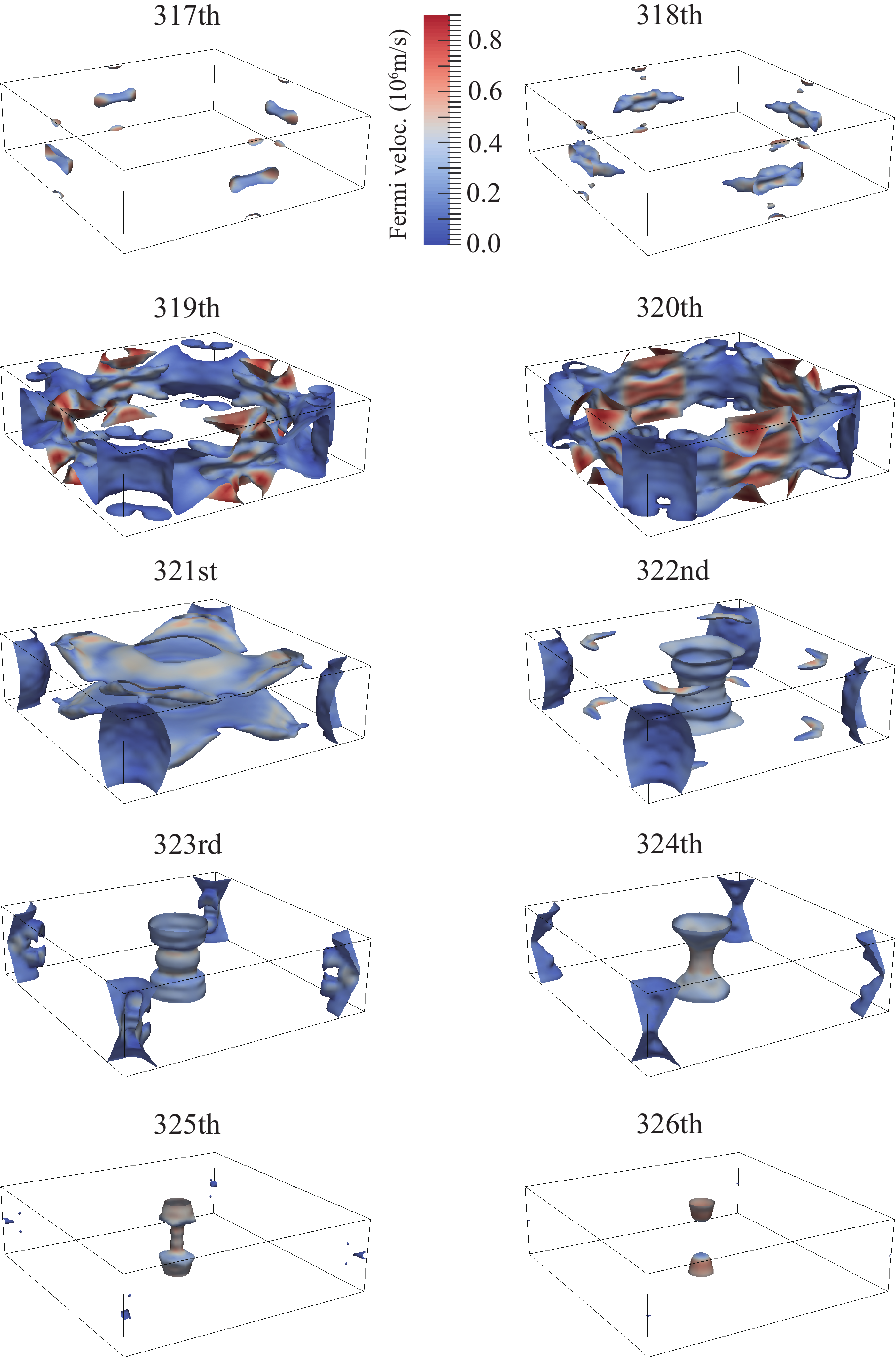}
	\end{center}
	\caption{\label{Fig8}(Color online) Calculated Fermi surface in the AF state. The numerical range of the color legend for the Fermi velocity is the same as that in Figs.~\ref{Fig6}(b) and \ref{Fig7}.}
\end{figure}

\subsection{Results and Discussion}
We performed calculations to self-consistently determine the charge density and the density matrices for the nonmagnetic PM and PPM states. For the AF state, Ce-$f^1$ occupation for the $|j=5/2, j_z=5/2\rangle$ orbital was assumed for the density matrices, taking the eight-CeRh$_2$Si$_2$-formula magnetic unit cell for the $4-{\bf q}$ magnetic order with the experimentally identified magnetic alignment.\cite{Kawarazaki2000} The PM Brillouin zone and the eight times smaller AF Brillouin zone are respectively shown in Figs.~\ref{Fig5}(a) and \ref{Fig5}(b). The different band structures and DOSs for the PM, PPM, and AF states are respectively shown in Figs.~\ref{Fig5}(c)-\ref{Fig5}(e). In the one-electron picture, the spin-orbit interaction splits the $f$ orbitals into $j=5/2$ and $j=7/2$ orbitals. The $j=7/2$ orbitals are located more than 1~eV above the Fermi level, and the large $U$ emphasizes the anisotropic occupation for the $j=5/2$ orbitals as discussed below.

The main result is that in the PM state the $f$ states located around the Fermi level contribute to the formation of the Fermi surface, in contrast to the $f$ orbitals located apart from the Fermi level in the PPM and AF states, which do not contribute to the Fermi surface close to the Fermi level.
In the PM state, $|j=5/2, j_z=\pm3/2\rangle$ orbitals mainly form the $f$ states around the Fermi level. As discussed in Refs. \citen{Suzuki2010} and \citen{Ikeda2015}, LDA calculations usually produce homogeneous $f$ states, to which all $j=5/2$ orbitals, meaning the three Kramers doublets $|j=5/2, j_z=\pm5/2\rangle, |j=5/2,j_z=\pm3/2\rangle, |j=5/2,j_z=\pm1/2\rangle$, contribute around the Fermi level because the small Coulomb potential $U$ is comparable to the $f$ bandwidth. By applying pressure, the crystal electrical field (CEF) anisotropy is suppressed, so the itinerant character of the $f$ electrons is more efficiently captured by performing LDA calculations above the critical pressure $p_c$ as discussed for CeCu$_2$Si$_2$.\cite{Ikeda2015}

In the AF state, the $|j=5/2, j_z=5/2\rangle$ orbitals are mainly occupied and are present at $-1$~eV below the Fermi level, while the $f$ states in the PPM state have a strong contribution from the $|j=5/2, j_z=\pm3/2 \rangle$ states mixed with the $|j=5/2, j_z=\pm5/2 \rangle$ states. The rich band structure of the AF state in Fig. \ref{Fig5}(e) is due to the band folding. The conduction bands of CeRh$_2$Si$_2$ originate from Si-$p$ and Rh-$d$ bands, where the latter form flat bands around the Fermi level, as seen in the band structures of the PPM and AF states in Figs. \ref{Fig5}(d) and \ref{Fig5}(e), respectively. These bands can contribute to the relatively large "band" masses of the Fermi surfaces even when the $f$ states do not contribute to the Fermi surfaces as shown in Figs.~\ref{Fig7} and \ref{Fig8}.  Although the lattice parameters can affect the electronic structure, the pressure or field effect only continuously modifies the conduction bands and hardly affects the $f$ electrons localized around the Ce atoms.

The corresponding Fermi surfaces in the PM, PPM and AF phases are shown in Figs.~\ref{Fig6}-\ref{Fig8}, respectively. In Fig. \ref{Fig6}(a), the $f$ orbital contribution for the Fermi surfaces is mapped on the Fermi surfaces. In the PM state, the $f$ orbital only has a large contribution on the 41$^{\rm st}$ Fermi surface as a result of the $k$-dependence of the hybridization between the $f$ states and conduction bands around the Fermi level. The Fermi velocity, which is defined as the $k$-space gradient of the energy bands, is also shown in Figs.~\ref{Fig6}-\ref{Fig8}. The blue color corresponds to a low Fermi velocity (large effective mass) and the red color to a large Fermi velocity (small effective mass). In the PM state, for the hole Fermi surface (band 40) the contribution of $f$ electrons is small, whereas the electron Fermi surface (band 41) has a large $4f$ contribution. The donut-like hole Fermi surface from band 40 has a similar shape to the band 25-hole Fermi surface obtained from the previous LDA calculations. \cite{note2, Araki2001, Harima} The only difference is the appearance of a hole in the donut in the LDA+$U$  calculation, which is absent in the LDA calculation. The shape of the heavy Fermi surface of band 41 is very different from that obtained from the previous calculation as it is strongly affected by the Coulomb repulsion $U$. These changes in the Fermi surfaces are caused by the different $k$-dependence of the hybridization between the $f$ states and conduction bands.

\begin{table}[h!!]
\caption{\label{Table2}Calculated quantum oscillation frequencies in the PM state of CeRh$_2$Si$_2$ for $H\parallel a$ and $H\parallel c$ obtained from LDA+$U$ (this work) and LDA (see Refs.~\citen{Araki2001} and \citen{Harima}) compared with those observed in dHvA experiments above $p_c$ (Ref.~\citen{Araki2001}).}
\centering
\begin{tabular}{cccc}
\hline
 &LDA+$U$ &  LDA \cite{Araki2001, Harima} & dHvA (13--16.9~T)\cite{Araki2001} \\
field & $F$ (T) (band) & $F$ (T) (band) & $F$ (T)\\
\hline
%\multicolumn{3}{c}{Frequencies (T) for $H\parallel a$ in the PM phase}\\
% \hline
%\multicolumn {1}{c}{ } & $F$~(T) & $m^\star$~($m_0$) & $F$~(T) & $m^\star$~($m_0$) &$F$~(T) & $m^\star$~($m_0$)   \\

$H \parallel a $ & 3400 (40$^{\rm th}$)& 9200 (26$^{\rm th}$)  & 8000  \\
 & 2500 (41$^{\rm th}$)& 4700 (25$^{\rm th}$)& 5800  \\
 & 1700 (41$^{\rm th}$)& 3400 (25$^{\rm th}$)& 4000\\
 & 1100 (40$^{\rm th}$)& 580   (27$^{\rm th}$)& \\
 \hline
% \multicolumn{3}{c}{Frequencies (T) for $H\parallel c$ in the PM phase}\\
% \hline
$H \parallel c $ &7766 (40$^{\rm th}$)&  10200 (26$^{\rm th}$)&\\
  & 905 (41$^{\rm st}$)&  9400 (25$^{\rm th}$)&\\
  & 748  (41$^{\rm st}$)& 330 (27$^{\rm th}$)&\\
  & 747 (41$^{\rm st}$)& &\\
  & 712(41$^{\rm st}$)& &\\
  & 710 (41$^{\rm st}$)& &\\
  & 648 (41$^{\rm st}$)& &\\
  & 196 (40$^{\rm th}$)& &\\
  \hline
\end{tabular}
\end{table}

The reason why small Fermi velocities persist in each phase can be found from the band structures around the Fermi level in Fig.~\ref{Fig5}, i.e., many of the energy bands have small dispersion around the Fermi level, leading to Fermi surfaces on which the band masses are relatively large, even in the AF and PPM phases. The LDA+$U$ calculations for the magnetic states show large splitting between the lower and upper Hubbard bands, which results in a weak contribution to the Fermi surfaces. These magnetic $f$states can be considered as localized in the sense that only conduction bands dominantly form the Fermi surfaces. As a result, the Fermi surfaces of the PPM state, which do not break translation symmetry with the localized $f$ states, are found to be similar to those calculated for LaRh$_2$Si$_2$.\cite{Araki2001} Bands 78 and 79 are hole surfaces and bands 82 and 83 are electron surfaces. Since CeRh$_2$Si$_2$ has an even number of total electrons, the Fermi surfaces in the PM state show the compensation of electron and hole Fermi surfaces. The FM polarization in the PPM phase breaks the compensation relation of the Fermi surfaces owing to the broken time-reversal symmetry. The sets  of Fermi surfaces with a very similar shape imply that the magnetic splitting of the energy bands around the Fermi level is very small, on the order of 0.1~eV at most, within the current calculations for the PPM phase.

In the AF state, the magnetic Brillouin zone, which is eight times smaller than that in the PM and PPM states, includes octuple band states due to the band folding (see Fig.~\ref{Fig5}), and these bands produce 20 Fermi surfaces with a degenerated pair due to the AF symmetry leading to a large number of dHvA branches (see Fig.~\ref{Fig8}). There is good agreement between the predicted frequencies in the AF2 state and the experimentally observed ones. However, the main frequency observed in the TEP experiment $\nu\sim 1770~$T was not predicted (see Table~\ref{Table}). The largest mass (6.4~$m_0$) observed in the dHvA measurements is associated with the largest frequency $a=7560~$T. This frequency may correspond to the largest orbit of the 319$^{\rm th}$ (7489~T) band originating from band folding. The calculated mass of this orbit is $m^*=2.26~m_0$. Note that this orbit has no donut-like shape as previously predicted.\cite{Araki2001} The largest mass observed by TEP measurements originates from the newly detected frequency at around 1370~T (with $m^*=7.2~m_0$). This frequency may come from one orbit of the 319$^{\rm th}$ band. The calculated effective mass of 1.8~$m_0$ associated with this frequency is considerably different from the observed value but it is still consistent with a large effective mass if we compare this value with the mass predicted for the $a$ branch. Indeed, the calculated masses are underestimated, which is due to the fact that the dynamical correlations are not taken into account in the LDA+$U$ calculation, so only qualitative comparison of the masses is appropriate. Note that the donut-like Fermi surface of the 40$^{\rm th}$ band in the PM state with a large frequency ($\sim 7700$~T) is similar to the  spin split Fermi surfaces (78$^{\rm th}$  and 79$^{\rm th}$ bands) with a large frequency ($\sim 9800$~T) in the PPM state. These Fermi surfaces are similar to those predicted and observed in a hole band of LaRh$_2$Si$_2$, \cite{Araki2001} and the similarities can be understood from the weak contribution of $f$ states of the 40$^{\rm th}$  Fermi surface as shown in Fig.~\ref{Fig6}. This means that the 40$^{\rm th}$ Fermi surface is dominantly formed by conduction bands.

Because quantum oscillations are detected for the ground state, comparison between the experiments and calculation in the PM phase is meaningful only above the critical pressure $p_c$ where the AF order collapses.\cite{structure} In previous dHvA measurements showing almost no pressure dependence in the AF state up to $p_c$, three large frequencies were detected above $p_c$ for $H\parallel a$ denoted $B\sim$4~kT, $A\sim5.8$~kT, and $C\sim8$~kT, indicating an abrupt change in the Fermi surfaces at $p_c$.\cite{Settai2003} The observed frequencies in the high pressure PM state are closer to the frequencies obtained in the previous calculation based on the LDA \cite{Araki2001} than those in the present calculation based on the LDA+$U$ (see Table~\ref{Table2}). This discrepancy can be explained by the fact that up on approaching $p_c$, the system may pass through a valence transition crossover.\cite{PalacioMorales2015} The influence of the crystal field collapses when the Kondo temperature exceeds the crystal field splitting, leading to a degenerated $4f$ electronic Kondo singlet and the suppression of the Ising character of the $4f$ electrons.\cite{Flouquet2009} In this context, as discussed for $p>p_c$, LDA calculations are more appropriate for calculating the band structure than LDA+$U$ calculations. The definitive test will be the observation of quantum oscillations for $H\parallel c$ in the PM regime. Unfortunately, no signal has yet been observed. The LDA calculations (see Fig.~10 from Ref.~\citen{Araki2001}) predict large frequencies originating from both the hole and electron Fermi surfaces, while the LDA+$U$ gives only a large hole and small electron orbits (see Fig.~\ref{Fig6}).

\section{Conclusion}

The Fermi surface of the AF phase of CeRh$_2$Si$_2$ was studied by examining its quantum oscillations in the TEP. Good agreement was observed with the new band structure calculations, which predict a large number of frequencies in the AF state due to band folding induced by the reduction of the Brillouin zone. The difference in the Fermi velocities between the PM and PPM or AF states is not clear, apart from the fact that the 4$f$ contribution changes from itinerant in the PM state to localized in the PPM or AF state, as predicted by a DOS calculation. The flatness of the band dispersion in the AF and PPM states contributes to the decrease in the Fermi velocity.

LDA+$U$ band structure calculations, which initially assumed that the $4f$ electrons were itinerant, stressed the contrasting behavior in the PM, AF, and PPM phases. In the PM phase the itinerant character of the $4f$ electrons appears directly on the Fermi surface topology. This is in excellent agreement with the high temperature observation that CeRh$_2$Si$_2$ is a heavy-fermion compound close to an AF magnetic instability.\cite{Willers2012,Flouquet2005a} However, to proceed from this qualitative comparison to a definite conclusion requires new attempts to obtain more information on the Fermi surface. A major breakthrough is expected from the observation of quantum oscillations for $H\parallel c$ above $p_c$.

The advanced treatment of the local correlations, for instance, by an LDA+DMFT (dynamical mean-field theory) approach, is expected to improve the present results and its application is a further issue. In the PM state at ambient presure, the strength of the AF correlations may push towards a localized picture above $T_N$. Angle-resolved photoemission spectroscopy (ARPES) measurements may clarify this challenge. In the PPM phase, the difficulty is the lack of extensive experimental data (only one frequency has been detected above $H_c$).\cite{Goetze2017} Even for the highly studied case of CeRu$_2$Si$_2$, a definitive comparison between the calculated Fermi surface topology and results of experiments remains limited. \cite{Julian1994, Aoki1995, Miyake2006, Daou2006, Boukahil2014} For example, the persistence of the itinerant character of $4f$ electrons may be hidden by the huge effective mass carried by the minority spin carriers as they will be subjected to the strong local repulsion of the majority spin sites. A key experimental challenge will be to obtain definitive microscopic insights on the Fermi surface in the PPM phase. There is now a wide range of heavy-fermion compounds where large magnetic polarization leads to Fermi surface instabilities, regardless of the nature of the AF correlations, of Ising type plus metamagnetism for CeRu$_2$Si$_2$ and CeRh$_2$Si$_2$, and of Heisenberg type for YbRh$_2$Si$_2$. In the latter case, in contrast to CeRh$_2$Si$_2$, far from a magnetic instability ($H>H_c$) the Fermi surface changes continuously under a magnetic field with successive Lifshitz transitions induced by an instability in the spin-polarized band.\cite{Pourret2013b, Pfau2013} Above $p_c$ the metamagnetism of CeRh$_2$Si$_2$ will be replaced by pseudo-metamagnetism, as intensively discussed for the CeRu$_2$Si$_2$ series.\cite{Flouquet2005a}
 The interplay between the $4f$ electron nature in heavy-fermion systems (localized or itinerant, Ising or Heisenberg; quasi-trivalent or mixed-valent) and magnetic polarization is the key ingredient to understand Fermi surface instabilities induced by magnetic field.

\acknowledgment{
We thank  H.~Harima, Y. \=Onuki, and S. Araki for many useful discussions. This work has been supported by the French ANR (PRINCESS project), the ERC (NewHeavyFermion starting grant), ICC-IMR, JSPS KAKENHI Grant Number 15H05883 (J-Physics), REIMEI, and EuromagNET II (EU contract no. 228043). LNCMI-CNRS is member of the European Magnetic Field Laboratory (EMFL).}

%\bibliographystyle{apsrev4-1}
%\bibliography{biblio}	

\begin{thebibliography}{26}
\bibitem{note} As the angular dependence of these new frequencies has not yet been determined, it is difficult to compare it with band structure calculations.
\bibitem{oscillations} In Fig.~\ref{Fig3} we show the magnetic field dependence of the quantum oscillation frequencies for the inverse field scale $\frac{1}{H_{\rm min}}-\frac{1}{H_{\rm max}}=0.03$~T$^{-1}$. Thus, $H_{\rm eff} = 18.2$~T for the highest field range. This field range was chosen to guarantee sufficient oscillations in the field range.  We also calculated FFT spectra for the high frequencies $\nu$, $\mu$, and $\epsilon$ for smaller inverse field ranges, and thus higher $H_{\rm eff}$ and observed no change in the field dependence.
\end{thebibliography}

%

\end{document}